\documentclass[fleqn,11pt]{article}
\usepackage{epsfig}
\input{amssym.def}
\input{amssym.tex}
\author{ }
\date{ }
\makeatletter

\def\gsim{\compoundrel>\over\sim}
\def\lsim{\compoundrel<\over\sim}
\def\compoundrel#1\over#2{\mathpalette\compoundreL{{#1}\over{#2}}}
\def\compoundreL#1#2{\compoundREL#1#2}
\def\compoundREL#1#2\over#3{\mathrel
      {\vcenter{\hbox{$\m@th\buildrel{#1#2}\over{#1#3}$}}}}
\begin{document}
%





\vspace*{2.0cm}
\begin{center}
{\Huge\bf An~SLC-type~$e^{+}e^{-}/\gamma\gamma$~facility}\\ 
\vspace*{0.6cm}
{\Huge\bf at~a~Future~Circular~Collider}\\
\vspace*{4.4cm}
{\Large R. BELUSEVIC}\\
\vspace*{0.8cm}
{\large {\em High Energy Accelerator Research Organization} (KEK)}\\
\vspace*{1.3mm}
{\large 1-1 {\em Oho, Tsukuba, Ibaraki} 305-0801, {\em Japan}} \\
\vspace*{1.3mm}
{\large belusev@post.kek.jp}\\
\end{center}

\thispagestyle{empty}

\newpage

\tableofcontents
\addtocontents{toc}{\protect\vspace{1.3cm}}
\vspace*{5mm}
\noindent
{\large\bf References}


\newpage

\vspace*{0.5cm}
\begin{center}
\begin{minipage}[t]{13.6cm}
{\bf Abstract\,:}\hspace*{3mm}
{It is proposed to place the arcs of an SLC-type facility inside the tunnel of a
Future Circular Collider (FCC). Accelerated by a linear accelerator (linac),
electron and positron beams would traverse the bending arcs in opposite
directions and collide at centre-of-mass energies considerably exceeding those
attainable at circular $e^{+}e^{-}$ colliders. The proposed SLC-type facility
would have the same luminosity as a conventional two-linac $e^{+}e^{-}$ collider.
Using an optical free-electron laser, the facility could be converted into a
$\gamma\gamma$ collider. A superconducting L-band linac at the proposed facility
may form a part of the injector chain for a 100-TeV proton collider in the FCC
tunnel. The whole accelerator complex would serve as a source of $e^{+}e^{-}$,
$\gamma\gamma$, $pp$ and $ep$  interactions. The L-band linac could also be used
to produce high-intensity neutrino, kaon and muon beams for fixed-target
experiments, as well as X-ray free-electron laser (XFEL) photons for applications
in material science and medicine.}
\end{minipage}
\end{center}

\vspace*{0.5cm}
\renewcommand{\thesection}{\arabic{section}}
\section{~Introduction}
\vspace*{0.3cm}

\setcounter{equation}{0}

~~~~The {\em Standard Model} (SM) of particle physics gives a coherent
quantum-mechanical description of electromagnetic, weak and strong interactions
based on fundamental constituents --- quarks and leptons --- interacting via
force carriers --- photons, W and Z bosons, and gluons. The SM is supported by
two theoretical `pillars': the {\em gauge principle} and the {\em Higgs
mechanism} for particle mass generation. In the SM, where electroweak symmetry is
broken by the Higgs mechanism, the mass of a particle depends on its interaction
with the Higgs field, a medium that permeates the universe. The photon and the
gluon do not have such couplings, and so they remain massless. The SM predicts
the existence of a neutral spin-0 particle associated with the Higgs field, but
it does not predict its mass.

Whereas the gauge principle has been firmly established through precision
electroweak measurements, the Higgs mechanism is yet to be fully tested. A state
decaying to several distinct final states was observed in 2012 at the CERN
Large Hadron Collider (LHC)  with a statistical significance of five standard
deviations \cite{ATLAS, CMS}. The observed state has a mass $\mbox{\large{$m$}}
_{\mbox{\tiny{H}}}^{~} \approx 125$ GeV. Its production rate is consistent with
the predicted rate for the SM Higgs boson. Furthermore, event yields in different
production topologies and different decay modes are self-consistent \cite{PDG}.

All of the couplings of the Higgs particle to gauge bosons and fermions are
completely determined in the SM in terms of electroweak coupling constants and
fermion masses. In the SM, Higgs production and decay processes can be computed
unambiguously in terms of the Higgs mass. Since the coupling of the Higgs boson
to fermions and gauge bosons is proportional to the particle masses, the Higgs
boson is produced in association with heavy particles and decays into the
heaviest particles that are kinematically accessible.

The Higgs-boson mass affects the values of electroweak observables through
radiative corrections. Many of the electroweak measurements obtained over the
past three decades may be combined to provide a global test of consistency with
the SM. The best constraint on $\mbox{\large{$m$}}_{\mbox{\tiny{H}}}^{~}$ is
obtained by making a global fit to the electroweak data. Such a fit strongly
suggests that the most likely mass for the SM Higgs boson is just above the limit
of 114.4 GeV set by direct searches at the LEP $e^{+}e^{-}$ collider \cite{LEP}.
This is consistent with the value of the Higgs mass measured at LHC.

High-precision electroweak measurements, therefore, provide a natural 
complement to direct studies of the Higgs sector. All the measurements made at
LEP and SLC could be repeated at the proposed facility using 90\% polarized
electron beams and at much higher luminosities \cite{erler}.

The rich set of final states in $e^{+}e^{-}$ and $\gamma\gamma$ collisions at the
proposed SLC-type facility would play an essential role in measuring the mass,
spin, parity, two-photon width and trilinear self-coupling of the SM Higgs boson,
as well as its couplings to fermions and gauge bosons. Such measurements require
centre-of-mass (c.m.) energies $\sqrt{s_{ee}} \lsim 600$ GeV, considerably
exceeding those attainable at circular $e^{+}e^{-}$ colliders.

\vspace*{0.3cm}
\section{~Single SM Higgs production in $e^{+}e^{-}$ annihilations}
\vspace*{0.3cm}

\begin{figure}[t]
\begin{center}
\epsfig{file=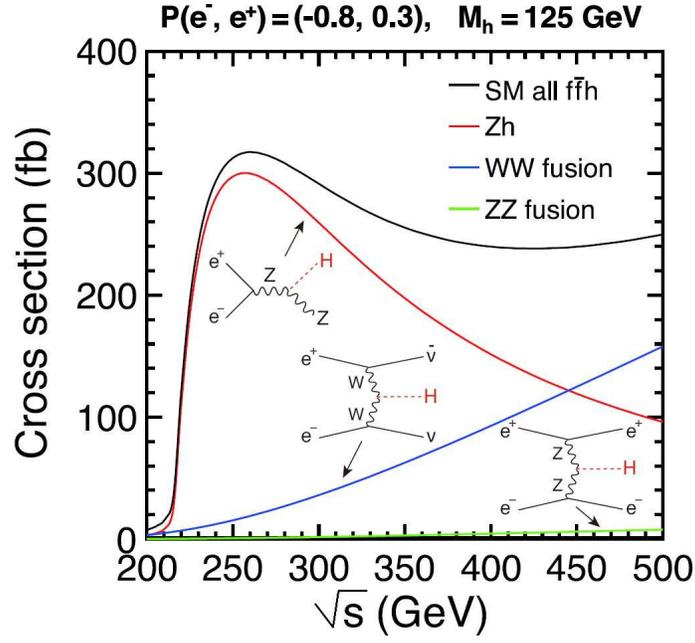,width=0.56\textwidth}
\end{center}
\vskip -7mm
\caption{Centre-of-mass energy dependence of the cross-sections for SM
Higgs-boson production in the Higgs-strahlung, W-fusion and Z-fusion processes
\cite{fujii0}.}
\label{fig:Xsections2}
\end{figure}

~~~~A particularly noteworthy feature of an $e^{+}e^{-}$ collider is that
the Higgs boson can be detected in the {\em Higgs-strahlung process}
(see Fig.\,\ref{fig:Xsections2})
\begin{equation}
e^{+}e^{-} \,\rightarrow\, {\rm HZ},\hspace*{3cm}\mbox{\large{$\sigma$}}(e^{+}
e^{-} \rightarrow {\rm HZ}) \,\propto\, \lambda_{\mbox{\tiny{HZZ}}}^{2}/s_{ee}
\end{equation}
even if it decays into invisible particles (e.g., the lightest {\em neutralino}
of a {\em supersymmetric model}). In this case the signal manifests itself as a
peak in the invariant mass distribution of the system which recoils against the 
lepton pair stemming from Z-boson decay. In Eq. (1), $\lambda_{\mbox{\tiny
{HZZ}}}$ is the Higgs coupling to the Z boson and $s_{ee}$ is the square of the
c.m. energy.

By exploiting the ${\rm HZ} \rightarrow X\ell^{+}\ell^{-}$ channel, the
Higgs-strahlung {\em cross-sections} can be measured with a statistical error of
about 2 percent for a Higgs-boson mass $\mbox{\large{$m$}}_{\mbox{\tiny{H}}}^{~}
\simeq 125$ GeV (see \cite{heinemeyer} and references therein). From the fits to
the reconstructed mass spectra in the channels ${\rm HZ} \rightarrow q\bar{q}
\ell^{+}\ell^{-},~b\bar{b}q\bar{q},~{\rm WW}\ell^{+}\ell^{-}$ and ${\rm WW}q
\bar{q}$, the {\em Higgs-boson mass} can be determined with an uncertainty of
about 40 MeV for $\mbox{\large{$m$}}_{\mbox{\tiny{H}}}^{~} \simeq 125$ GeV
\cite{heinemeyer}.

To determine the {\em spin} and {\em parity} of the SM Higgs boson
in the Higgs-strahlung process, one can use the information on (1) the energy
dependence of the Higgs-boson production cross-section just above the kinematic
threshold, and (2) the angular distribution of the Z/H bosons. The best way to
study the {\em CP properties} of the Higgs boson is by analyzing the spin
correlation effects in the decay channel ${\rm H} \rightarrow \tau^{+}\tau^{-}$
\cite{heinemeyer}.

\begin{figure}[!h]
\begin{center}
\epsfig{file=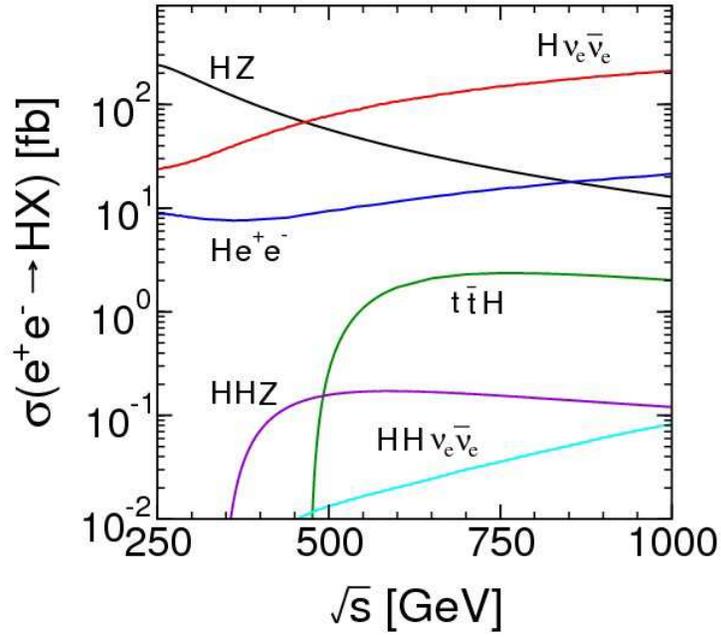,width=0.59\textwidth}
\end{center}
\vskip -7mm
\caption{Centre-of-mass energy dependence of various cross-sections for single and double SM Higgs-boson production in $e^{+}e^{-}$ annihilations
\cite{asner0}.}
\label{fig:Xsections3}
\end{figure}

The Higgs-strahlung cross-section, which dominates at low c.m. energies,
decreases with energy in proportion to $1/s$. In contrast, the cross-section for
the {\em W-fusion process} (see Figs \ref{fig:Xsections2} and
\ref{fig:Xsections3})
\begin{equation}
e^{+}e^{-} \,\rightarrow\, {\rm H}\nu\bar{\nu},\hspace*{3cm}
\mbox{\large{$\sigma$}}(e^{+}e^{-} \rightarrow {\rm H}\nu\bar{\nu}) \,\propto\,
\lambda_{\mbox{\tiny{HWW}}}^{2}\log\!\mbox{\Large{$($}}s_{ee}/\mbox{\large{$m$}}_
{\mbox{\tiny{H}}}^{\,2}\mbox{\Large{$)$}}
\end{equation}
increases with energy in proportion to log($s_{ee}/\mbox{\large{$m$}}_{\mbox
{\tiny{H}}}^{\,2}$), and hence becomes more important at energies $\sqrt{s_{ee}}
\gsim 500$ GeV for $\mbox{\large{$m$}}_{\mbox{\tiny{H}}}^{~} \simeq 125$ GeV. In
Eq. (2), $\lambda_{\mbox{\tiny{HWW}}}$ is the Higgs coupling to the W boson.

The Higgs-fermion couplings can be extracted by measuring the {\em branching 
fractions} of the Higgs boson. There are two methods to determine the Higgs
branching fractions: (1) Measure the event rate in the Higgs-strahlung process
for a given final-state configuration and then divide by the total cross-section;
(2) Select a sample of unbiased events in the Higgs-strahlung recoil-mass peak
and determine the fraction of events that correspond to a particular decay
channel. See \cite{heinemeyer} and references therein for an estimate of the
accuracy that can be achieved in such measurements.

For $\mbox{\large{$m$}}_{\mbox{\tiny{H}}}^{~} < 2\mbox{\large{$m$}}_{\mbox
{\tiny{W}}}^{~}$, the {\em total decay width} of the Higgs boson, $\Gamma
_{\mbox{\tiny{H}}}$, can be determined indirectly by employing the relation
between the total and partial decay widths for a given final state:
\begin{equation}
\Gamma_{\mbox{\tiny{H}}} \,=\, \frac{\Gamma ({\rm H} \rightarrow X)}
{{\rm BR}({\rm H} \rightarrow X)}
\end{equation}
For instance, consider the decay ${\rm H} \rightarrow {\rm WW}^{*}$. One can
directly measure the branching fraction ${\rm BR}({\rm H} \rightarrow
{\rm WW}^{*})$, determine the coupling HZZ in the process $e^{+}e^{-} \rightarrow
{\rm HZ}$, relate the HZZ and HWW couplings ($\lambda_{\mbox{\tiny{HZZ}}}^{~}/
\lambda_{\mbox{\tiny{HWW}}}^{~} =  \mbox{\large{$m$}}_{\mbox{\tiny{Z}}}^{2}/2
\mbox{\large{$m$}}_{\mbox{\tiny{W}}}^{2}$), and then use the fact that $\Gamma
({\rm H} \rightarrow {\rm WW}) \propto \lambda_{\mbox{\tiny{HWW}}}^{2}$ to obtain
the partial width $\Gamma ({\rm H} \rightarrow {\rm WW}^{*})$ from the
information on the HWW coupling. The accuracy with which the determination of
$\Gamma_{\mbox{\tiny{H}}}$ can be achieved for $\mbox{\large{$m$}}_{\mbox
{\tiny{H}}}^{~} \simeq 125$ GeV is estimated in \cite{heinemeyer}.

\newpage

\vspace*{1.8cm}

\newpage

\renewcommand{\thesection}{\arabic{section}}
\section{~Single SM and MSSM Higgs production in $\gamma\gamma$ collisions}
\vspace*{0.3cm}

~~~~Since photons couple directly to all fundamental fields carrying the
electromagnetic current (leptons, quarks, W bosons, supersymmetric particles),
$\gamma\gamma$ collisions provide a comprehensive means of exploring virtually
every aspect of the SM and its extensions (see \cite{boos, belusev} and
references therein). The cross-sections for production of charged-particle pairs
in $\gamma\gamma$ interactions are approximately an order of magnitude larger
than in $e^{+}e^{-}$ annihilations. For some processes within and beyond the SM,
the required c.m. energy is considerably lower in $\gamma\gamma$ collisions than
in $e^{+}e^{-}$ or proton-proton interactions.

In $\gamma\gamma$ collisions, the Higgs boson is produced as a {\em single
resonance} in a state of definite CP, which is perhaps the most important
advantage over $e^{+}e^{-}$ annihilations, where this $s$-channel process is
highly suppressed. At c.m. energies $\sqrt{s_{ee}} \lsim 500$ GeV, the effective
cross-section for
\begin{equation}
\gamma\gamma \,\rightarrow\, {\rm H}
\end{equation}
is at least a factor of four larger than any cross-section for Higgs production
in $e^{+}e^{-}$ annihilations. Moreover, the process $e^{+}e^{-}\rightarrow
{\rm HZ}$ requires considerably higher c.m. energies than $\gamma\gamma
\rightarrow$ H.

Any theoretical model based on the gauge principle must evoke spontaneous
symmetry breaking. In the {\em minimal supersymmetric} extension of the Standard
Model ({\small MSSM}), for instance, spontaneous electroweak symmetry breaking
results in five physical Higgs-boson states: two neutral scalar fields $h^0$ and
$H^0$, a pseudoscalar $A^0$ and two charged bosons $H^\pm$. In $e^{+}e^{-}$
annihilations, the heavy neutral MSSM Higgs bosons can be created only by
associated production ($e^{+}e^{-} \rightarrow H^{0}A^{0}$), whereas in
$\gamma\gamma$ collisions they are produced as single resonances 
($\gamma\gamma \rightarrow H^{0},\,A^{0}$) with masses up to 80\% of the initial
$e^{-}e^{-}$ collider energy \cite{zerwas}. For example, if their masses are
around 500 GeV, then $H^{0}$ and $A^{0}$ could be produced either in pairs in
$e^{+}e^{-}$ annihilations at $\sqrt{s_{ee}} \simeq 1$ TeV, or as single particles in $\gamma\gamma$ collisions at $\sqrt{s_{ee}} \sim 600$ GeV.

The reaction $\gamma\gamma \rightarrow$ H, which is related to ${\rm H}
\rightarrow \gamma\gamma$, proceeds through a `loop diagram' and receives
contributions from {\em all} charged particles that couple to the photon and
the Higgs boson. Thus, the {\em two-photon width} $\Gamma ({\rm H}
\rightarrow \gamma\gamma )$ is sensitive to the Higgs-top Yukawa coupling, as
well as mass scales far beyond the energy of the $\gamma\gamma$ collision.
Assuming that the branching ratio ${\rm BR}({\rm H} \rightarrow b\bar{b})$ can
be measured to an accuracy of about 2\% in the process $e^{+}e^{-} \rightarrow
{\rm HZ}$, the $\gamma\gamma$ partial width can be determined with a similar 
precision by measuring the cross-section
\begin{equation}
\mbox{\large{$\sigma$}}(\gamma\gamma \rightarrow
{\rm H} \rightarrow b\bar{b}) \propto \Gamma ({\rm H} \rightarrow \gamma\gamma)
{\rm BR}({\rm H} \rightarrow b\bar{b})
\end{equation}
Each of the decay modes ${\rm H} \rightarrow \bar{b}b$,\,WW can be measured in
photon-photon collisions with a precision comparable to that expected from
analyses based on $e^{+}e^{-}$ data (see, e.g., \cite{asner}).

High-energy photons can be produced by Compton-backscattering of laser light on
electron beams. Both the energy spectrum and polarization of the backscattered 
photons depend strongly on the polarizations of the incident electrons and
laser photons. The key advantage of using $e^{-}e^{-}$ beams is that they can
be polarized to a high degree, enabling one to tailor the photon energy 
distribution to one's needs. In a $\gamma\gamma$ collision, the possible
helicities are 0 or 2. The Higgs boson is produced in the $J_{z} = 0$ state,
whereas the background processes $\gamma\gamma \rightarrow b\bar{b},\,c\bar{c}$
are suppressed for this helicity configuration. The circular polarization of the
photon beams is therefore an important asset, for it can be used both to enhance
the signal and suppress the background.

The CP {\em properties} of any neutral Higgs boson that may be produced 
at a photon collider can be {\em directly} determined by controlling the 
polarizations of Compton-scattered photons \cite{grzadkowski}. A CP-even Higgs
boson couples to the combination ${\bf e}_{\mbox{\tiny{1}}}\mbox{\boldmath
{$\cdot$}}\,{\bf e}_{\mbox{\tiny{2}}}$, whereas a CP-odd Higgs boson couples to
$({\bf e}_{\mbox{\tiny{1}}}\!\times\!{\bf e}_{\mbox{\tiny{2}}})\,\mbox{\boldmath
{$\cdot$}}\,\mbox{\boldmath{$k$}}_{\gamma}$, where ${\bf e}_{i}^{~}$ are
polarization vectors of colliding photons, $\phi$ is the angle between them, and
$\mbox{\boldmath{$k$}}_{\gamma}$ is the momentum vector of one of the
Compton-scattered photons. The scalar (pseudoscalar) Higgs boson  couples to
{\em linearly polarized} photons with a maximum strength if the polarization
vectors are parallel (perpendicular). 

The general amplitude for a CP-{\em mixed state} to couple to two photons can be
expressed as 
\begin{equation}
{\cal M} \,=\, {\cal E}({\bf e}_{\mbox{\tiny{1}}}\mbox{\boldmath{$\cdot$}}\,
{\bf e}_{\mbox{\tiny{2}}}) + {\cal O}({\bf e}_{\mbox{\tiny{1}}}\!\times\!{\bf e}
_{\mbox{\tiny{2}}})_{z}^{~}
\end{equation}
where ${\cal E}$ is the CP-even and ${\cal O}$ the CP-odd contribution to the
amplitude. If we denote the {\em helicities} of the two photons by $\lambda_{1}$
and $\lambda_{2}$, with $\lambda_{1},\lambda_{2} = \pm 1$, then the above vector
products read
${\bf e}_{\mbox{\tiny{1}}}\mbox{\boldmath{$\cdot$}}\,{\bf e}_{\mbox{\tiny{2}}} =
-(1 + \lambda_{1}\lambda_{2})/2$ and $({\bf e}_{\mbox{\tiny{1}}}\!\times\!{\bf e}
_{\mbox{\tiny{2}}})_{z}^{~} = i\lambda_{1}(1 + \lambda_{1}\lambda_{2})/2$.
As shown in \cite{grzadkowski}, one can define three {\em polarization
asymmetries} that yield an unambiguous measure of CP-mixing. Note that it is
necessary to have both {\em linearly} and {\em circularly} polarized photons in
order to measure those asymmetries. In $e^{+}e^{-}$ annihilations, it is possible
to discriminate between CP-even and CP-odd neutral Higgs bosons, but would be
difficult to detect small CP-violating effects (which contribute only at the
one-loop level) for a dominantly CP-even component (which contributes at the tree
level in  $e^{+}e^{-}$ collisions) \cite{hagiwara}.

A study of single Higgs-boson production in $\gamma\gamma$ collisions via the
hadronic content of the photon ({\em resolved processes}) was reported in
\cite{doncheski}. Such contributions to $\gamma\gamma \rightarrow$ H are
non-negligible. Resolved photon production of the heavy MSSM Higgs bosons
$H^{0}$ and $A^{0}$ would complement other measurements by probing particular regions of the SUSY parameter space \cite{doncheski}.

To ascertain the physics potential of a $\gamma\gamma$ collider, one must take
into account the fact that the photons are not monochromatic \cite{ginzburg}. As
already mentioned, both the energy spectrum and polarization of the backscattered
photons depend strongly on the polarizations of the incident electrons and
photons. A longitudinal electron-beam polarization of 90\% and a 100\% circular
polarization of laser photons are customarily assumed.

\vspace*{0.3cm}
\section{~Higgs-pair production in $\gamma\gamma$ and $e^+e^-$ collisions}
\vspace*{0.3cm}

~~~~It is well known that hadron colliders are not ideally suited for measuring
the self-coupling of the Higgs boson if $\mbox{\large{$m$}}_{\mbox{\tiny{H}}}
^{~}\,\leq\,140$~GeV \cite{baur}. The potential of a future $\gamma\gamma /
e^+e^-$ collider for determining the HHH coupling has therefore been closely 
examined (see \cite{belusev1} and \cite{Djouadi:1999gv}--\cite{ belusev2}).

\begin{figure}[t]
\begin{center}
\epsfig{file=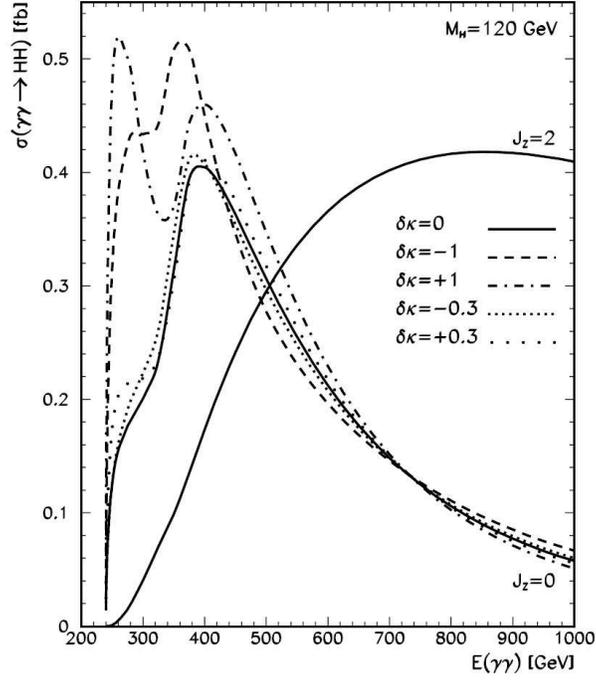,width=0.48\textwidth}
\end{center}
\vskip -7mm
\caption{The cross-sections for HH production in $\gamma\gamma$ collisions for
$\mbox{\large{$m$}}_{\mbox{\tiny{H}}}^{~}\,=\,120$~GeV and anomalous trilinear
Higgs self-couplings $\delta\kappa = 0,\,\pm 0.3,\,\pm 1$. Credit: R. Belusevic
and J. Jikia \cite{belusev1}.}
\label{fig:Higgs_self}
\end{figure}

The production of a pair of SM Higgs bosons in photon-photon collisions,
\begin{equation}
\gamma\gamma \,\to\, {\rm HH}
\end{equation}
which is related to the Higgs-boson decay into two photons, is due to W-boson and
top-quark box and triangle loop diagrams. The total cross-section for
$\gamma\gamma\to{\rm HH}$ in polarized photon-photon collisions, calculated at
the leading one-loop order \cite{Jikia:1992mt} as a function of the
$\gamma\gamma$ c.m. energy and for $\mbox{\large{$m$}}_{\mbox{\tiny{H}}}^{~}$
between 115 and 150 GeV, is given in \cite{belusev1}. The cross-section
calculated for equal photon helicities, $\mbox{\large{$\sigma$}}_
{\gamma\gamma\,\rightarrow\,\mbox{\tiny{HH}}}(\mbox{\small{$J_{z}=0$}})$, rises
sharply above the $2\mbox{\large{$m$}}_{\mbox{\tiny{H}}}^{~}$ threshold for
different values of $\mbox{\large{$m$}}_{\mbox{\tiny{H}}}^{~}$, and has a peak
value of about $0.4$~fb at a $\gamma\gamma$ c.m. energy of 400~GeV. In contrast,
the cross-section for opposite photon helicities, $\mbox{\large{$\sigma$}}
_{\gamma\gamma\,\rightarrow\,\mbox{\tiny{HH}}}(\mbox{\small{$J_{z}=2$}})$, rises
more slowly with energy because a pair of Higgs bosons is produced in a state
with orbital angular momentum of at least $2\hbar$
(see Fig.\,\ref{fig:Higgs_self}).

The cross-sections for equal photon helicities are of special interest, since
only the $J_z=0$ amplitudes contain contributions with trilinear Higgs 
self-coupling. By adding to the SM Higgs potential $V(\Phi^{\dag}\Phi )$ 
a gauge-invariant dimension-6 operator $\mbox{\large{$($}}\Phi^{\dag}\Phi\mbox
{\large{$)$}}^{3}$, one introduces a gauge-invariant anomalous trilinear
Higgs coupling $\delta\kappa$ \cite{Jikia:1992mt}. For the reaction $\gamma
\gamma \to {\rm HH}$, the only effect of such a coupling in the {\em unitary
gauge} would be to replace the trilinear Higgs coupling of the SM,
$\lambda_{\mbox{\tiny{HHH}}}^{~}$, by an {\em anomalous Higgs self-coupling}
$\lambda \,=\, (1 + \delta\kappa )\lambda_{\mbox{\tiny{HHH}}}^{~}$. The
dimensionless anomalous coupling $\delta\kappa$ is normalized so that 
$\delta\kappa=-1$ exactly cancels the SM HHH coupling. The cross-sections
$\mbox{\large{$\sigma$}}_{\gamma\gamma\,\rightarrow\,\mbox{\tiny{HH}}}$ for
five values of $\delta\kappa$ are shown in Fig.\,\ref{fig:Higgs_self}.

In an experiment to measure the trilinear Higgs self-coupling, the contribution
from $\gamma\gamma \to {\rm HH}$ for opposite photon helicities represents
an irreducible background. However, this background is suppressed if one
chooses a $\gamma\gamma$ c.m. energy below about 320 GeV. 

The Feynman diagrams for the process $\gamma\gamma \rightarrow {\rm HH}$ are
shown in \cite{Jikia:1992mt}. New physics beyond the SM introduces additional
complexity into the subtle interplay between the Higgs `pole amplitudes' and the
top-quark and W-boson `box diagrams':
\begin{equation}
|{\cal M}(J_{z} = 0)|^{2} \,=\, |A(s)(\lambda_{\mbox{\tiny{SM}}}^{~} +
\delta\lambda ) \,+\, B|^{2}
\end{equation}
where $\lambda_{\mbox{\tiny{SM}}}^{~}$ is the trilinear Higgs self-coupling in
the SM. From this expression we infer that the cross-section
\begin{equation}
\mbox{\large{$\sigma$}}(\gamma\gamma \rightarrow {\rm HH}) \,=\, \alpha
\lambda^{2} + \beta\lambda + \gamma~~~~~~~~~~~~~~~\alpha > 0,~~\gamma > 0
\end{equation}
is a quadratic function of the coupling $\lambda \equiv \lambda_{\mbox{\tiny
{SM}}}^{~} + \delta\lambda$.

\begin{figure}[!h]
\begin{center}
\epsfig{file=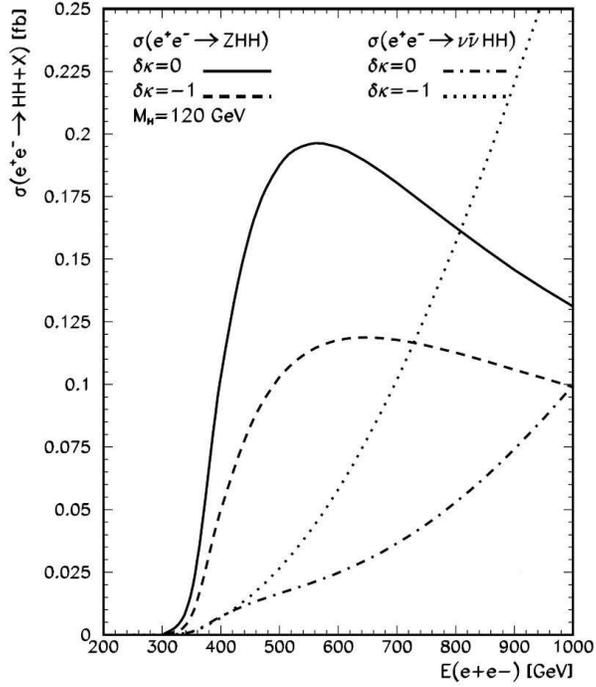,width=0.48\textwidth}
\end{center}
\vskip -7mm
\caption{The total cross-sections for $e^+e^- \rightarrow {\rm HHZ}$ and
$e^+e^- \rightarrow {\rm HH}\nu_e^{~}\bar{\nu}_e^{~}$ as functions of the
$e^+e^-$ c.m. energy for $\mbox{\large{$m$}}_{\mbox{\tiny{H}}}^{~}\,=\,120$~GeV
and anomalous trilinear Higgs self-couplings $\delta\kappa=0,\,\pm 0.3,\,\pm 1$.
Credit: R. Belusevic and J. Jikia \cite{belusev1}.}
\label{fig:Higgs_self1}
\end{figure}

The trilinear self-coupling of the Higgs boson can also be measured either in 
the so-called {\em double Higgs-strahlung process}
\begin{equation}
e^+e^- \,\to\, {\rm HHZ}
\end{equation}
or in the {\em W-fusion reaction}
\begin{equation}
e^+e^- \,\to\, {\rm HH}\nu_e^{~}\bar{\nu}_e^{~}
\end{equation}
The total cross-section for pair production of 120-GeV Higgs bosons in $e^+e^-$
collisions, calculated for {\em unpolarized} beams, are shown in
Fig.\,\ref{fig:Higgs_self1} for anomalous trilinear Higgs self-couplings $\delta
\kappa = 0$ or $-1$. If the electron beam is 100\% polarized, the double
Higgs-strahlung cross-section will stay approximately the same, while the
W-fusion cross-section will be twice as large. From the plots in
Fig.\,\ref{fig:Higgs_self1} we infer that the  SM double Higgs-strahlung
cross-section exceeds 0.1~fb at 400~GeV for $\mbox{\large{$m$}}_{\mbox{\tiny{H}}}
^{~}\,=\,120$~GeV, and reaches a broad maximum of about 0.2~fb at a c.m. energy
of 550~GeV. The SM cross-section for W-fusion stays below 0.1~fb for c.m.
energies up to 1 TeV.

For $\mbox{\large{$m$}}_{\mbox{\tiny{H}}}^{~}\,=\,120$~GeV, and assuming a
longitudinal electron-beam polarization of 90\%, the maximum sensitivity to an
anomalous trilinear Higgs self-coupling is achieved in the so-called double 
Higgs-strahlung process at a c.m. energy of about 500~GeV \cite{belusev1}. This
is significantly higher than the optimal c.m. energy in $\gamma\gamma$
collisions. In the W-fusion process, a similar sensitivity is attained at c.m.
energies well above 500 GeV.

Calculations show that the {\em statistical sensitivity} of
$\mbox{\large{$\sigma$}}_{\gamma\gamma\,\rightarrow\,\mbox{\tiny{HH}}}$ to the
Higgs self-coupling is maximal near the kinematic threshold for Higgs-pair
production if $\mbox{\large{$m$}}_{\mbox{\tiny{H}}} \sim 120$ GeV,
and is comparable with the sensitivities of $\mbox{\large{$\sigma$}}_{e^{+}
e^{-}\,\rightarrow\,\mbox{\tiny{HHZ}}}$ and $\mbox{\large{$\sigma$}}_{e^{+}
e^{-}\,\rightarrow\,\mbox{\tiny{HH}}\nu\bar{\nu}}$ to this coupling for 
$\sqrt{s_{ee}}\leq 700$ GeV, even if the integrated luminosity in $\gamma\gamma$
collisions is only one third of that in $e^+e^-$ annihilations \cite{belusev1}.
The overall {\em acceptance} should, in principle, be considerably larger in the
process $\gamma\gamma \rightarrow {\rm HH}$ than in the reaction $e^{+}e^{-} 
\rightarrow {\rm HHZ}$ due to the smaller final-state particle multiplicity.

Since the cross-section
$\mbox{\large{$\sigma$}}_{\gamma\gamma\,\rightarrow\,\mbox{\tiny{HH}}}$ does
not exceed 0.4 fb, it is essential to attain the highest possible luminosity,
rather than energy, in order to measure the trilinear Higgs self-coupling.
As shown in \cite{belusev1}, appropriate angular and invariant-mass cuts and a
reliable $b$-tagging algorithm are needed in order to suppress the dominant WW,
ZZ and four-quark backgrounds well below the HH signal.

The results of detailed feasibility studies for measuring Higgs-pair production
in $\gamma\gamma$ and $e^+e^-$ collisions have been reported \cite{kawada, tian}.
It has been shown that the optimum $\gamma\gamma$ collision energy is around 270
GeV for a 120-GeV Higgs boson, and that the main backrounds at this energy are
the processes $\gamma\gamma \rightarrow$ WW, ZZ and $b\bar{b}b\bar{b}$. The
preliminary analysis described in \cite{kawada} suggests that the process
$\gamma\gamma \rightarrow$ HH could be observed with a statistical significance
of about 5$\sigma$, provided proper color-singlet clustering is used in jet
reconstruction. The precision with which the trilinear Higgs self-coupling could
be measured in the process $e^+e^- \to {\rm HHZ}$ at $\sqrt{s_{ee}} = 500$ GeV
and in the reaction $e^+e^- \to {\rm HH}\nu_e^{~}\bar{\nu}_e^{~}$ at
$\sqrt{s_{ee}} = 1$ TeV is presented in \cite{tian}.

\newpage

\vspace*{0.7cm}

\newpage

\section{~The proposed facility}
\vspace*{0.3cm}

\begin{figure}[h]
\begin{center}
\epsfig{file=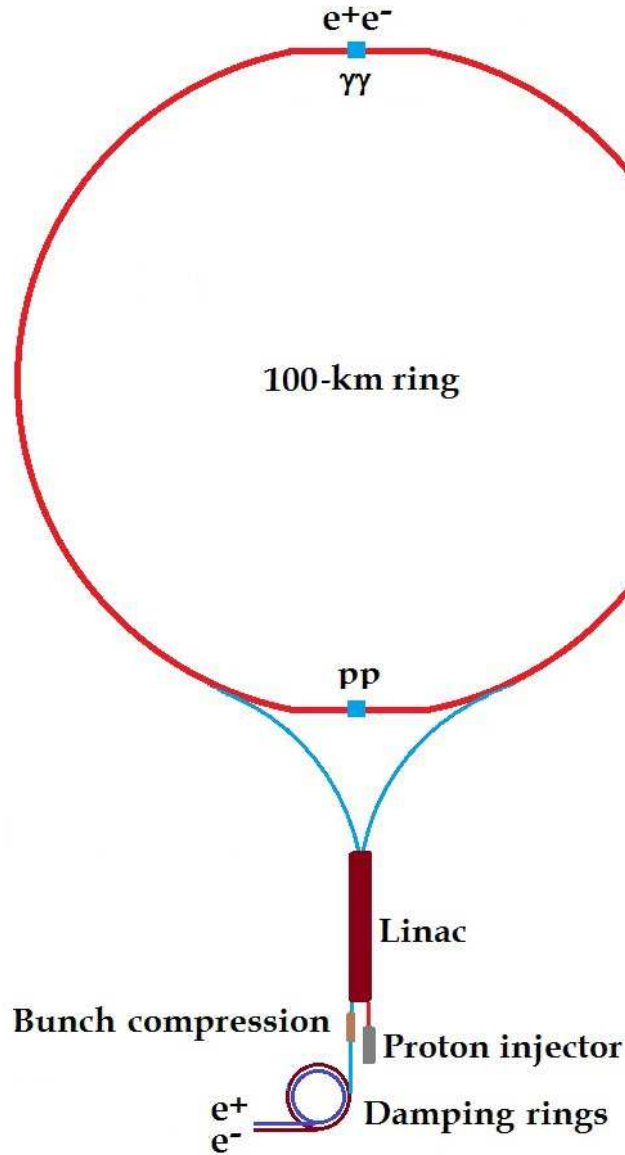,width=0.62\textwidth}
\end{center}
\vskip -7mm
\caption{Schematic layout of the proposed SLC-type facility. A 350-GeV
superconducting linac (with a focusing quadrupole in each cryomodule) could also
be a  part of the FCC injector chain.}
\label{fig:SLC}
\end{figure}

~~~~A schematic layout of the proposed SLC-type $e^{+}e^{-}/\gamma\gamma$
facility is shown in Fig.\,\ref{fig:SLC}. Damped and bunch-compressed electron
and positron beams, accelerated by a single linear accelerator (linac), traverse
two arcs of bending magnets in opposite directions and collide at an interaction
point surrounded by a detector. The beams are then disposed of, and this machine
cycle is repeated at a rate that depends on whether the linac is made of L-band
or X-band accelerating structures. Using an optical free electron laser (FEL),
high-energy photons for a $\gamma\gamma$ collider are created by Compton
backscattering of FEL photons on electrons prior to their collision.

With a crossing angle at the interaction point (IP), separate beam lines may be
used to bring the disrupted beams to their respective dumps, thereby enabling
post-IP diagnostics. It is also envisaged that a `bypass line' for low-energy
beams would be employed to accumulate data at the Z resonance in the process
$e^{+}e^{-}\rightarrow{\rm Z}$. These runs could be used to regularly calibrate
the detector, fine-tune the accelerator and measure its luminosity.

The proposed facility could be constructed in several stages, each with distinct
physics objectives that require particular center-of-mass (c.m.) energies:
\[
       \begin{array}{ll}
\bullet~~e^{+}e^{-} \rightarrow {\rm Z,\,WW};\hspace*{0.5cm}\gamma\gamma 
\rightarrow {\rm H}~~~~~~~&~~~~~~~\sqrt{s_{ee}} \sim 90~{\rm to}~180~{\rm GeV} 
\\*[4mm]
\bullet~~e^{+}e^{-} \rightarrow {\rm HZ}~~~~~~~&~~~~~~~\sqrt{s_{ee}} \sim 250~
{\rm GeV} \\*[4mm]
\bullet~~e^{+}e^{-} \rightarrow t\bar{t};\hspace*{0.5cm}\gamma\gamma \rightarrow
{\rm HH}~~~~~~~&~~~~~~~\sqrt{s_{ee}} \sim 350~{\rm GeV} \\*[4mm]
\bullet~~e^{+}e^{-}\rightarrow {\rm HHZ},\,{\rm H}t\bar{t},\,{\rm H}\nu\bar{\nu}
~~~~~~~&~~~~~~~\sqrt{s_{ee}} \gsim 500~{\rm GeV}
       \end{array} \]
For instance, the top-quark mass could be measured in the process $e^{+}e^{-}
\rightarrow t\bar{t}$ at the pair-production threshold; one expects
$\delta\mbox{\large{$m$}}_{t}^{~} \approx 100~{\rm MeV} \approx 0.1\delta\mbox
{\large{$m$}}_{t}^{~}({\rm LHC})$ \cite{fujii}.

The linac at the proposed SLC-type facility would consist either of (1)
high-gradient X-band cavities developed for CLIC and a corresponding
klystron-based power source (a two-beam scheme could be implemented at a later stage); or (2) ILC-type superconducting L-band cavities placed within cryogenic
vessels and fed by multi-beam klystrons.

The 11.4 GHz X-band rf technology was originally developed at SLAC and KEK. The
choice of this technology is motivated by the cost benefits of having relatively
low rf energy per pulse and high accelerating gradients. A comprehensive review
of the status of X-band accelerator technology is given in \cite{adolphsen}.
Since then, significant advances have been made in pulsed HV and rf power
generation, high gradient acceleration and wakefield suppression. The ultimate
design of rf cavities will depend on the outcome of the ongoing effort to develop
100 MeV/m X-band structures for a CLIC-type linear collider. 

As proposed in \cite{belusev2}, a single X-band rf unit contains a modulator that
drives a pair of 50 MW klystrons, each of which generates 1.6 $\mu$s rf pulses at
50 Hz. An rf compression system enhances the peak power of the klystrons by a
factor of 3.75, and produces 245 ns pulses that match the accelerator structure
requirements. The resulting 375 MW, 245 ns pulses feed seven 0.21m-long
accelerator structures, producing a 85 (100) MV/m loaded (unloaded) gradient in
each structure.

The current design for the {\em International Linear Collider} (ILC), based on
the superconducting technology originally developed at DESY, uses L-band
(1.3 GHz) superconducting niobium rf cavities that have average accelerating
gradients of 31.5 MeV/m (see \cite{ILC} and references therein). Nine cavities,
each 1 m long, are mounted together in a string and assembled into a common
low-temperature cryostat or {\em cryomodule}. Liquid helium is used to cool
cavities to $-271^{\circ}$ C.

An ILC-type main linac is composed of rf units, each of which is formed by three 
contiguous cryomodules containing 26 nine-cell cavities. Every unit has an rf
source, which includes a pulse modulator, a 10 MW multi-beam klystron, and a
waveguide system that distributes the power to the cavities. An ILC-type design
offers some advantages over the X-band technology:

\vspace*{1mm}
$\bullet$~~Wakefields are drastically reduced due to the large size of the rf
cavities, which means that cavity alignment tolerances can be relaxed. This is crucial for an SLC-type facility, where both $e^{+}$ and $e^{-}$ bunches are
alternately accelerated;

$\bullet$~~Superconducting rf cavities can be loaded using a long rf pulse
(1.5 ms) from a source with low peak rf power;

$\bullet$~~`Wall-plug to beam' power transfer efficiency is about twice that
of X-band cavities; 

$\bullet$~~The long rf pulse allows a long bunch train ($\sim 1$ ms), with many
bunches ($\sim 3000$) and a relatively large bunch spacing ($\sim 300$ ns). A
trajectory correction (feedback) system within the train can therefore be used
to bring the beams into collision.
\vspace*{1mm}

However, in contrast to a compact, high-gradient X-band machine, a collider 
based on the current ILC-type design would be characterized by ($a$) low
accelerating gradients; ($b$) two large damping rings with a total length of at
least six kilometers, and ($c$) a technologically challenging cryogenic system
that requires a number of surface cryogenic plants.

An important feature of the proposed SLC-type facility is the
possibility of using backscattered laser beams to produce high-energy
$\gamma\gamma$ collisions \cite{ginzburg}. In order to attain maximum luminosity
at a $\gamma\gamma$ collider, every electron bunch in the accelerator should
collide with a laser pulse of sufficient intensity for $63\%$ of the electrons to
undergo a Compton scattering. This requires a laser system with high average
power, capable of producing pulses that would match the temporal spacing of
electron bunches \cite{belusev2}.

These requirements could be satisfied by an optical {\em free electron laser} 
(FEL) \cite{saldin}. The radiation produced by an FEL has a variable wavelength,
and is fully polarized either circularly or linearly depending on whether the
undulator is helical or planar, respectively. The wavelength $\lambda$ of FEL
radiation is determined by $\lambda \approx \lambda_{u}/2\gamma^{2}$, where
$\gamma \equiv {\rm E}/m_{e}c^{2}$ is the {\em Lorentz factor} of the electron
beam with energy E and $\lambda_{u}$ is the periodic length of the undulator. To
produce photon pulses of required intensity, suitable high-intensity,
low-emittance rf guns have to be developed \cite{michelato}.

Assuming that the mean number of Compton interactions of an electron in a laser
pulse (the Compton conversion probability) is 1, the {\em conversion
coefficient} $k \equiv n_{\gamma}^{~}/n_{e} \approx 1 - \mbox{\small{e}}^{-1} =
0.63$, where $n_{e}$ is the number of electrons in a 'bunch' and $n_{\gamma}
^{~}$ is the number of scattered photons. The luminosity of a gamma-gamma
collider is then
\begin{equation}
{\cal L}_{\gamma\gamma} \,=\, (n_{\gamma}^{~}/n_{e}^{~})^{2\,}{\cal L}_{ee}
 \,\approx\, (0.63)^{2\,}{\cal L}_{ee}
\end{equation}
where ${\cal L}_{ee}$ is the {\em geometric luminosity} at a conventional
two-linac collider:
\begin{equation}
{\cal L}_{ee} \,\propto\, \frac{\gamma n_{e}^{\,2}N_{b}f}{\sqrt{\mbox{\large
{$\varepsilon$}}_{xn}\beta_{x}\mbox{\large{$\varepsilon$}}_{yn}\beta_{y}}} 
\,\equiv\, \frac{P_{\rm beam}}{\sqrt{s_{ee}}}\frac{\gamma n_{e}^{~}}{\sqrt{\mbox
{\large{$\varepsilon$}}_{xn}\beta_{x}\mbox{\large{$\varepsilon$}}_{yn}\beta_{y}}}
\end{equation}
In this expression, $\beta_{x}, \beta_{y}$ are the horizontal and vertical
{\em beta functions}, respectively, $\mbox{\large{$\varepsilon$}}_{xn},\mbox
{\large{$\varepsilon$}}_{yn}$ are the normalized transverse {\em beam
emittances}, $N_{b}$ is the number of bunches per rf pulse, $f$ is the pulse
{\em repetition rate}, $\sqrt{s_{ee}}$ is the c.m. energy, and $P_{\rm beam} =
n_{e}N_{b}f\sqrt{s_{ee}}$ is the {\em beam power}.

There are $N_{b}/2$ electron or positron bunches in each arc of an SLC-type
facility. If its repetition rate is twice that of a conventional two-linac
collider, so that roughly the same wall-plug power is used, the two machines
would have the same luminosity (see Eq. (13)).

The {\em energy loss} per turn due to {\em synchrotron radiation} (SR) in a
storage ring is given by
\begin{equation}
\Delta{\rm E} \,=\, C_{\gamma}\frac{\raisebox{-.4ex}{${\rm E}_{0}^{4}$}}
{\raisebox{.4ex}{$\rho$}}\hspace*{1.2cm}\Rightarrow\hspace*{1.2cm}{\rm E}(s)
\,=\, {\rm E}_{0}\!\left (1 \,+\, \frac{\raisebox{-.4ex}{$A$}}{\rho^{2}}s\right )
^{\!-1/3}
\end{equation}
where $C_{\gamma} = 88.46\times 10^{-6}$~m/GeV$^{3}$, E\,[GeV] is the beam energy, $\rho$\,[m] is the effective bending radius, $s$\,[m] is the beam path
length, and $A \equiv 3C_{\gamma}{\rm E}_{0}^{3}/2\pi$. For ${\rm E}_{0} = 250$
GeV and $\rho = 12$ km, the expression on the left yields $\Delta{\rm E} = 14.4$
GeV per half turn; for ${\rm E}_{0} = 350$ GeV and the same radius, $\Delta
{\rm E} = 55.3$ GeV per half turn. If there are no accelerating structures in the
arcs, the linac energy must be increased, e.g., from ${\rm E}_{0} = 300$ GeV to
${\rm E}_{0} \approx 350$ GeV in order to attain $\sqrt{s_{ee}} = 600$ GeV.

\begin{figure}[t]
\begin{center}
\epsfig{file=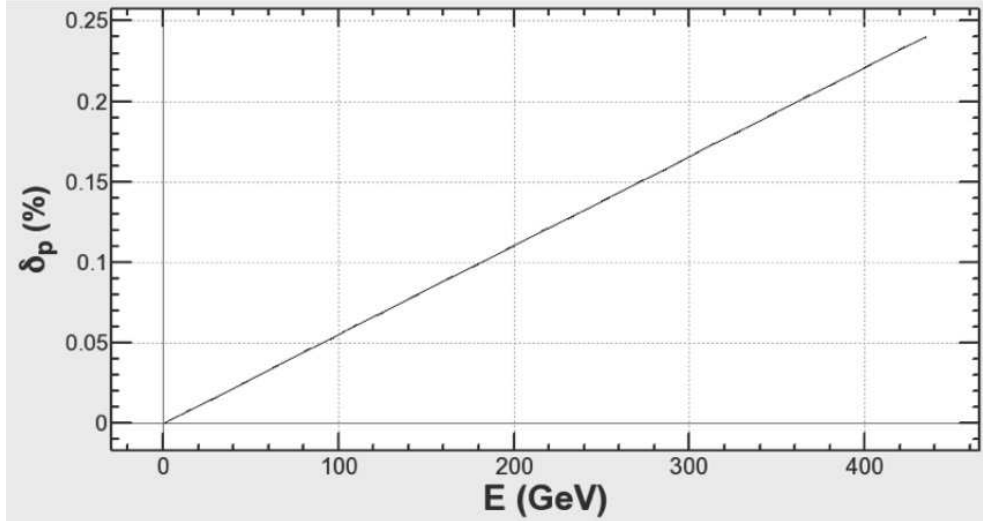,width=0.80\textwidth}
\end{center}
\vskip -7mm
\caption{Energy spread in an electron beam traversing an arc with an effective
bending radius $\rho = 12$ km. To produce this plot, a lattice of
combined-function FODO cells was used as an input to K. Oide's SAD tracking code.
Credit: D. Zhou, KEK.}
\label{fig:Espread}
\end{figure}

\begin{figure}[!h]
\begin{center}
\epsfig{file=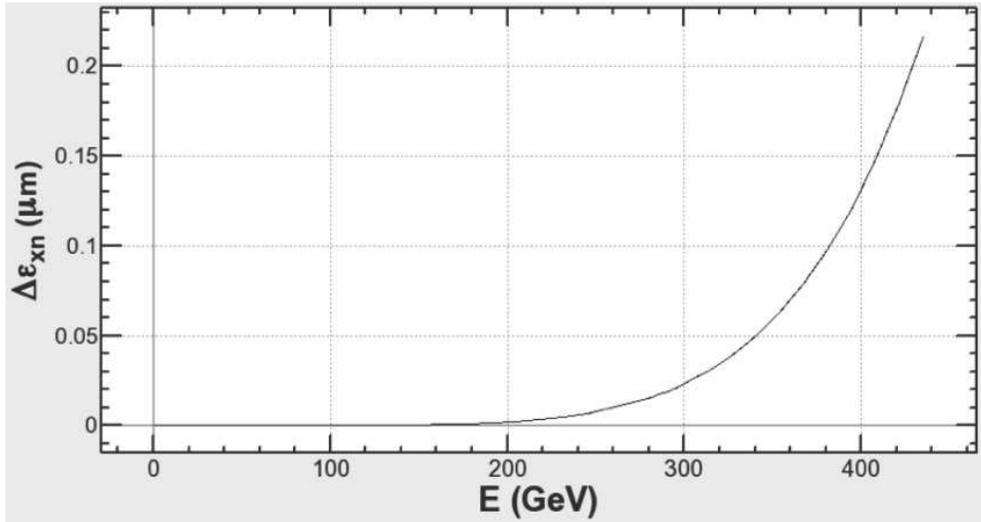,width=0.80\textwidth}
\end{center}
\vskip -7mm
\caption{Energy dependence of the growth of the horizontal electron beam
emittance in an arc with an effective bending radius $\rho = 12$ km. To produce
this plot, a lattice of combined-function FODO cells was used as an input to
K. Oide's SAD tracking code. Credit: D. Zhou, KEK.}
 \label{fig:emittance}
\end{figure}

The {\em critical energy} of SR photons, ${\rm E}_{c}\,{\rm [keV]} = 2.22\,
{\rm E}^{3}\,{\rm [GeV]}/\rho\,{\rm [m]}$, is approximately 8 MeV for
${\rm E} = 350$ GeV and $\rho = 12$ km. The {\em energy spread} in an
electron beam due to SR is given by
\begin{equation}
\frac{\mbox{\large{$\sigma$}}_{\rm E}^{~}}{\rm E} \,\approx\, \gamma\sqrt{\frac{C_{q}}{2\rho}}
\end{equation}
where $\gamma$ is the Lorentz factor of the beam, $C_{q}\approx 3.84\times
10^{-13}$ m, and $\rho$\,[m] is the bending radius. For ${\rm E} = 250$ GeV and
$\rho = 12$ km, Eq. (15) yields $\mbox{\large{$\sigma$}}_{\rm E}^{~}/{\rm E}
\approx 6\times 10^{-4}$ (cf. Fig.\,\ref{fig:Espread}). For ${\rm E} \lsim 450$
GeV, a preliminary calculation indicates that the growth of the {\em horizontal
beam emittance} in the bending arcs would not exceed the value $\mbox{\large
{$\varepsilon$}}_{xn} = 2\,\mu$m at KEK's ATF damping ring (see
Fig.\,\ref{fig:emittance}).

In contrast to ILC or CLIC, an SLC-type collider would have a single bunch
compression system and a short beam transfer line connecting the damping rings
with the entrance to the main linac (see Fig.\,\ref{fig:SLC}). A 350-GeV
superconducting L-band linac at the proposed facility may form, together with a
3-TeV energy booster, the injector chain for a proton collider in the FCC
tunnel (e.g., the linac could replace the chain LINAC4 $\rightarrow$ PSB
$\rightarrow$ PS $\rightarrow$ SPS at CERN).

\newpage

\vspace*{3cm}

\newpage

\section{~Concluding remarks}
\vspace*{0.3cm}

~~~~It is proposed to place the arcs of an SLC-type facility inside the 100 km long tunnel of a Future Circular Collider (FCC). Electron and positron beams,
accelerated in a single X-band or L-band linac, would traverse the arcs of
bending magnets in opposite directions (see Fig.\,\ref{fig:SLC}) and collide at
c.m. energies considerably exceeding those attainable at circular $e^{+}e^{-}$
colliders. Using an optical free-electron laser (FEL), the SLC-type facility
could be converted into a $\gamma\gamma$ collider. Large savings in
construction cost could be achieved if the crossing angle and the beam dump are
exactly the same for the operation of the SLC-type facility in the $e^{+}e^{-}$
and $\gamma\gamma$ collision modes.

The proposed $e^{+}e^{-}/\gamma\gamma$ collider could be built in several stages,
each with distinct physics objectives that require particular c.m. energies (see
Sections 2--5). The following unique features of the proposed facility are
particularly noteworthy:

\vspace*{3mm}
$\bullet$~~The maximum luminosity at a circular $e^{+}e^{-}$ collider is severely
constrained by beamstrahlung effects at high energies; also, it is very difficult
to achieve a high degree of beam polarization \cite{koratzinos}. This is not the
case at an SLC-type facility, where luminosity is proportional to beam energy and
the electron beam polarization can reach about 90\%. The availability of
polarized beams is essential for some important precision measurements in
$e^{+}e^{-}$ and $\gamma\gamma$ collisions \cite{moortgat}.

\vspace*{1mm}
$\bullet$~~It is straightforward to convert an SLC-type facility into a
high-luminosity $\gamma\gamma$ collider with highly polarized beams. This
considerably increases its physics potential (see below).

\vspace*{1mm}
$\bullet$~~A 350-GeV superconducting L-band linac at the proposed facility may
form, together with a 3-TeV energy booster, the injector chain for a 100-TeV
proton collider in the FCC tunnel. The L-band linac could also be used to produce
high-intensity neutrino, kaon and muon beams for fixed-target experiments, as
well as X-ray FEL photons for applications in material science and medicine
\cite{radoje}.

\vspace*{1mm}
$\bullet$~~If electron or positron bunches, accelerated by the L-band linac at
the proposed facility, are brought into collision with the 50-TeV FCC proton
beams, the whole accelerator complex could serve also as a source of
deep-inelastic $ep$ interactions \cite{akay}. Such interactions would yield
valuable information on the quark-gluon content of the proton, which is crucial
for precision measurements at the FCC hadron collider . The physics potential of
an $ep$ collider is discussed, e.g., in \cite{LH_ep}.
\vspace*{3mm}

The rich set of final states in $e^{+}e^{-}$ and $\gamma\gamma$ collisions at the
proposed SLC-type facility would play an essential role in measuring the mass,
spin, parity, two-photon width and trilinear self-coupling of the SM Higgs boson,
as well as its couplings to fermions and gauge bosons. Such measurements require
c.m. energies considerably exceeding those attainable at circular $e^{+}e^{-}$
colliders. For instance, one has to measure separately the couplings HWW, HHH and
Htt at $\sqrt{s_{ee}} \gsim 500$ GeV in order to determine the corresponding SM
loop contributions to the effective HZZ coupling (see Sections 2--5 and, e.g.,
\cite{McCullough}).

For some processes within and beyond the SM, the required c.m. energy is
considerably lower in $\gamma\gamma$ collisions than in $e^{+}e^{-}$ or
proton-proton interactions. For example, the heavy neutral MSSM Higgs bosons can
be created in $e^{+}e^{-}$ annihilations only by associated production
($e^{+}e^{-} \rightarrow H^{0}A^{0}$),  whereas in $\gamma\gamma$ collisions they
are produced as single resonances ($\gamma\gamma \rightarrow H^{0},\,A^{0}$) with
masses up to 80\% of the initial $e^{-}e^{-}$ collider energy.

Both the energy spectrum and polarization of the backscattered photons at a
$\gamma\gamma$ collider depend strongly on the polarizations of the incident
electrons and laser photons. The circular polarization of the photon beams
is an important asset, for it can be used both to enhance the signal and suppress
the background. The CP properties of any neutral Higgs boson produced at a photon
collider can be directly determined by controlling the polarizations of
Compton-scattered photons.

\newpage

\section{~Acknowledgements}
\vspace*{0.3cm}

~~~~I would like to thank I. Ginzburg, T. Higo, A. Wolski and K. Yokoya for
valuable comments and suggestions concerning various aspects of this proposal.
I am especially grateful to K. Oide and D. Zhou for helping me estimate some
relevant beam properties at the proposed facility.

\end{document}